\documentclass[fleqn,usenatbib,useAMS]{mnras}
\usepackage{amssymb,amsmath}
\usepackage{graphicx}
\usepackage{xcolor,natbib}
\usepackage{multirow}
\usepackage[T1]{fontenc}
\usepackage{ae,aecompl}
\usepackage{newtxtext, newtxmath}
\usepackage{float}
\usepackage{subfigure}
\usepackage{longtable}
\usepackage{enumitem}
\usepackage{longtable,listings}
\usepackage[flushleft]{threeparttable}
\usepackage{parcolumns}
\def\dif{\mathop{}\!\mathrm{d}}
\def\obj{SDSS J0752}

\title[6.4yr QPOs in SDSS J0752]{A 6.4yr optical quasi-periodic oscillations in SDSS J075217.84+193542.2: 
a new candidate for central binary black hole system}

\author[Zhang X. G.]{XueGuang Zhang$^{1}$
\thanks{Contact e-mail: \href{mailto:xgzhang@njnu.edu.cn}{xgzhang@njnu.edu.cn}}\\
$^{1}$School of Physics and technology, Nanjing Normal University,
          No. 1, Wenyuan Road, Nanjing, 210023, P. R. China}

\begin{document}
\label{firstpage}
\pagerange{\pageref{firstpage}--\pageref{lastpage}}

\maketitle

\begin{abstract} 
	In this manuscript, a 6.4yr optical quasi-periodic oscillations (QPOs) is detected in the 
quasar SDSS J075217.84+193542.2 (=\obj) at a redshift 0.117, of which 13.6yr-long light 
curve from CSS and ASAS-SN directly described by a sinusoidal function with a periodicity 6.4yr. 
The 6.4yr QPOs can be further confirmed through the Generalized Lomb-Scargle periodogram with 
confidence level higher than 99.99\%, and through the auto-correlation analysis results, and through 
the WWZ technique. The optical QPOs strongly indicate a central binary black hole (BBH) system in 
\obj. The determined two broad Gaussian components in the broad H$\alpha$ can lead to the BBH system 
with expected space separation about 0.02pc between the expected two central BHs with determined 
virial BH masses about $8.8\times10^7{\rm M_\odot}$ and $1.04\times10^9{\rm M_\odot}$. Meanwhile, 
we check the disk precessions applied to explain the optical QPOs. However, under the disk precession 
assumption, the determined optical emission regions from central BH have sizes about $40{\rm R_G}$ 
two times smaller than sizes of the expected NUV emission regions through the correlation between 
disk size and BH mass, indicating the disk precessions are not preferred. And due to the lower radio 
loudness around 0.28, jet precessions can be also totally ruled out. Furthermore, only 
0.08\% probability can determined as the QPOs mis-detected through light curves randomly created 
by the CAR process, re-confirming the reported optical QPOs.
\end{abstract}

\begin{keywords}
galaxies:active - galaxies:nuclei - quasars:emission lines - galaxies:Seyfert
\end{keywords}

\section{Introduction}

	Optical Quasi-Periodic Oscillations (QPOs) with long periodicities of years to more than 
ten years have been reported in active galaxies, due to jet emissions/precessions in blazars as 
discussed in \citet{sc18, bg19, oa20} or due to binary supermassive BH (BBH) systems as discussed 
in \citep{eh94, kf09, gm10, bb15, sx20}. Not similar as transient QPOs arising from general 
relativistic effects (relativistic Frame Dragging method \citep{ref1}, discoseismology method 
\citep{ref2}, etc.) related to central accreting processes in black hole X-ray binaries as 
discussed in \citet{vm98, vm00, ak04, rm06, jp10, vi17, im20} and in several Active Galactic Nuclei 
(AGN) as discussed in \citet{pl93, mk06, gm08, li13, ps14, rr16, dl17, bs18, sm18, gt18, jd21, 
zh21a}, the QPOs related to BBH systems are due to orbital motions of two BH accreting systems 
leading to stronger QPOs signals which can be directly detected in the long-term light curves. 
In the manuscript, unless otherwise stated, the following discussed optical QPOs are the optical 
QPOs related to BBH systems.

	Evolution from dual galactic core systems on scale of dozens to hundreds parsecs to 
BBH systems on scale of sub-parsecs have been accepted to be common, as the key role of galaxy 
mergers in evolution histories of galaxies \citep{bb80, sr98, md06, mk10, ml13, mb19}. And different 
techniques have been applied to detect dual core systems with the two black holes getting closer 
due to dynamical friction and/or BBH systems with two black holes getting closer due to emission 
of gravitational waves, such as based on double-peaked features of broad and/or narrow emission 
lines as discussed in \citet{zw04, kz08, bl09, ss09, sl10, eb12, pl12, cs13, le16, wg17, dv19} 
and on spatially resolved image properties of central regions of galaxies as discussed in \citet{km03, 
rt09, pv10, ne17, kw20} and on apparent QPOs signals detected in the long-term variability properties 
which are mainly considered in the manuscript.

     PG 1302-102 is the well-known quasar with apparent periodic variabilities with a periodicity 
1800days reported in \citet{gd15a}, providing strong evidence for the central BBH system in PG 
1302-102. More recent results on the optical QPOs in PG 1302-102 can be found in \citet{lg18, kp19}. 
\citet{gd15} have searched for strong Keplerian periodic signals over a baseline of nine years and 
reported a sample of 111 candidates with variabilities are conservative agreement with theoretical 
predictions from BBH systems. \citet{cb16} have conducted a statistical search for periodic 
variabilities in a sample of 35383 spectroscopically confirmed quasars, and identified 50 quasars 
with significant QPOs with periodicities of a few hundred days. \citet{lg15} have detected strong 
QPOs with a periodicity about 540days in the quasar PSO J334.2028+01.4075, leading to a reliable 
candidate for the central BBH system. And \citet{zb16} have shown the clear QPOs with a periodicity 
about 1500days in the quasar SDSS J0159+0105, supporting a reliable BBH system. More recently, 
\citet{ss20} have detected the apparent QPOs with a periodicity about 1150days in the Seyfert1.5 
Mrk 915, providing robust clues on the central BBH system. \citet{ky20} have reported 
the clear QPOs with periodicity about 1.2yr in the Mrk 231 related to the central BBH system. 
\citet{lw21} have reported a QPOs with periodicity about 1607days with confidence level higher 
than 99.95\% in the quasar SDSS J025214.67-002813.7 to support a central BBH system. 


	The reported candidates of BBH systems should produce expected background gravitational 
wave signals at nano-Hz frequencies, probed by the Pulsar Timing Arrays \citep{fb90, de16, re16, 
ar15, ve16}. However, for the candidates of BBH systems in \citet{gd15, cb16}, \citet{se18} have 
shown that the null hypothesis (whereby the candidates of BBH systems are false positives) is preferred 
over the binary hypothesis at about 2.3$\sigma$ and 3.6$\sigma$ for the candidates in \citet{gd15} and 
in \citet{cb16} respectively, indicating the current candidates of BBH systems have some false 
candidates due to false QPOs detections. Therefore, to detect and report more candidates of BBH 
systems is necessary and meaningful. And in the manuscript, a new candidate of BBH system is reported.

	Optical QPOs related to BBH systems, not similar as weak transient QPOs, are commonly strong 
enough that the QPOs can be directly detected from the long-term light curves of AGN, even considering 
intrinsic AGN variabilities as fundamental characteristics of AGN \citep{mr84, um97, ms16, bg20} which 
have been proved to well described by the well-applied Continuous AutoRegressive process (CAR process 
or the improved damped random walk process (DRW process)) \citep{kbs09, koz10, zk13, kb14, sh16, zk16, 
zh17}. In this manuscript, a new BBH candidate is reported in the blue quasar SDSS J075217.84+193542.2 
(=\obj) at a redshift 0.117 \citep{pa18}, due to detected optical QPOs through the long-term 
variabilities from the Catalina Sky Survey (CSS) \citep{dr09} and from the All-Sky Automated Survey for 
Supernovae (ASAS-SN) \citep{sp14, ks17}. The shown time duration of the \obj is two times longer than 
the detected periodicity, indicating the optical QPOs in \obj should be robust to some extent. 
Moreover, the time duration from 2005 to 2013\ in CSS of the \obj~ is only about a little more than one 
cycle (considering the following determined 6.4year periodicity), leading the \obj~ not included in the 
sample of candidates of BBH systems in \citet{gd15} with observing time durations in CSS larger than 1.5 
cycles.

	The manuscript is organized as follows. Section 2 presents main results on the long-term optical 
variabilities of \obj. Section 3 shows main results on the spectroscopic properties of SDSS J0752. Section 
4 gives the necessary discussions on the probable central BBH system. Section 5 gives final summaries and 
conclusions. Here, the cosmological parameters have been adopted as $H_{0}=70{\rm km\cdot s}^{-1}{\rm Mpc}^{-1}$, 
$\Omega_{\Lambda}=0.7$ and $\Omega_{\rm m}=0.3$.

\begin{figure*}
\centering\includegraphics[width = 18cm,height=6cm]{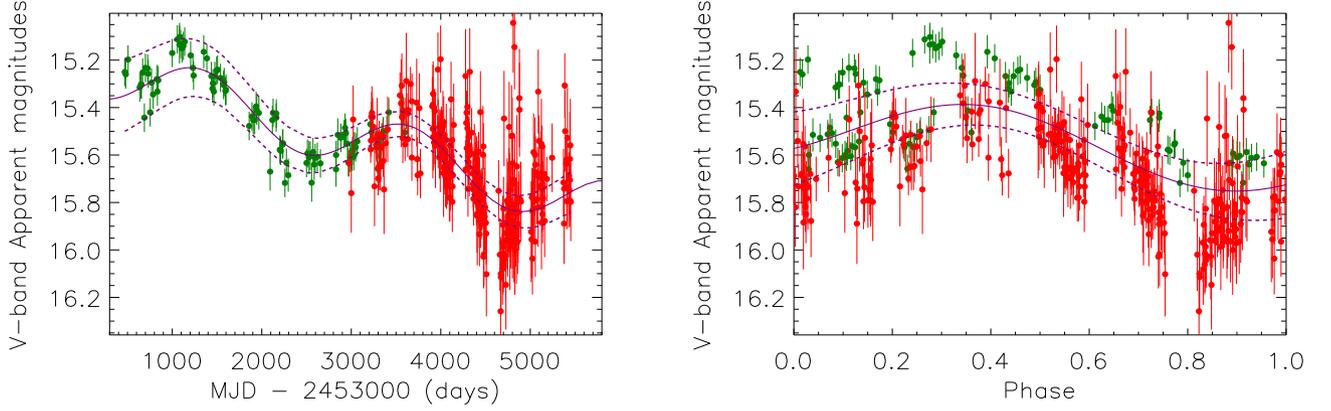}
\caption{Left panel shows the long-term light curve from the CSS (in dark green) and from the ASAS-SN 
(in red color). Right panel shows the corresponding phase folded light curve based on the determined 
periodicity of 2321days. In each panel, solid and dashed lines in purple show the best descriptions to 
the whole light curve and the corresponding 99.9999\% confidence bands, based on a sinusoidal 
function plus a linear trend. 
}
\label{lmc}
\end{figure*}

\section{Long-term optical variabilities in SDSS J0752}

\subsection{Optical QPOs in SDSS J0752}

	The CSS V-band light curve is collected from \url{http://nesssi.cacr.caltech.edu/DataRelease/} with 
MJD-2453000 from 469 (April 2005) to 3596 (October 2013), and the ASAS-SN V-band light curve is collected 
from \url{https://asas-sn.osu.edu/} with MJD-2453000 from 2974 (February 2012) to 5452 (November 2018). Left 
panel of Figure~\ref{lmc} shows the long-term photometric V-band light curve. Here, we do not show the ASAS-SN 
g-band data points in Figure~\ref{lmc}, because of the unknown magnitude difference between ASAS-SN V-band 
and ASAS-SN g-band and of the short time duration of the ASAS-SN g-band light curve. And moreover, similar 
as what have been done to the light curves from the CSS and from the ASAS-SN in the PG 1302-102 discussed in 
\citet{lg18}, the magnitude difference 0.2019 is accepted, based on the data points in time duration with 
MJD-2453000 from 2973 to 3597 covered both in the CSS and in the ASAS-SN.

	Then, by both Levenberg-Marquardt least-squares minimization technique and Maximum Likelihood 
Method combining with Markov Chain Monte Carlo (MCMC) technique \citep{fh13}, a simple sinusoidal function 
plus a linear trend are applied to determine the best descriptions to the 13.6yr-long light curve, with 
$\chi^2/Dof\sim1.28$ ($\chi^2$ and Dof as the summed squared residuals and the degree of freedom). Here, 
the main objective of applications of sinusoidal function is to show clearer clues on optical QPOs. 
The best-fitting results are shown in left panel of Figure~\ref{lmc} by a linear trend plus 
a sinusoidal function 
\begin{equation}
LMC~=~A~+~B~\times~\frac{t}{\rm 1000days}~+~C~\times~\sin(\frac{2\pi t}{T_{QPOs}}~+~\phi_0)
\end{equation}
with $A~=~15.223\pm0.012$, $B~=~0.103\pm0.003$, $C~=~0.119\pm0.007$, $T_{QPOs}~=~2321\pm32$, 
$\phi_0~=~1.159\pm0.131$, leading to the QPOs with a periodicity of $2321\pm32$ days.
Moreover, based on the F-test technique, the corresponding 99.9999\% confidence bands to the best fitting 
results are shown in left panel of Figure~\ref{lmc}. Then, based on the determined periodicity, the phase 
folded light curve $LMC_{pf}$ shown in the right panel of Figure~\ref{lmc} can also be well described by 
a sinusoidal function plus a linear trend
\begin{equation}
LMC_{pf}~=~A~+~B~\times~\frac{t}{\rm 1000days}~+~C~\times~\sin(2\pi t~+~\phi_0)
\end{equation}
with $A~=~15.475\pm0.013$, $B~=~0.152\pm0.026$, $C~=~-0.142\pm0.008$, $\phi_0~=~-0.753\pm0.061$. 
And based on the F-test technique, the corresponding 99.9999\% confidence bands to the best fitting results 
are also shown in right panel of Figure~\ref{lmc}. The directly fitted results by the sinusoidal function 
to both the light curve and the phase folded light curve strongly support the optical QPOs in SDSS J0752.

\begin{figure*}
\centering\includegraphics[width = 18cm,height=6cm]{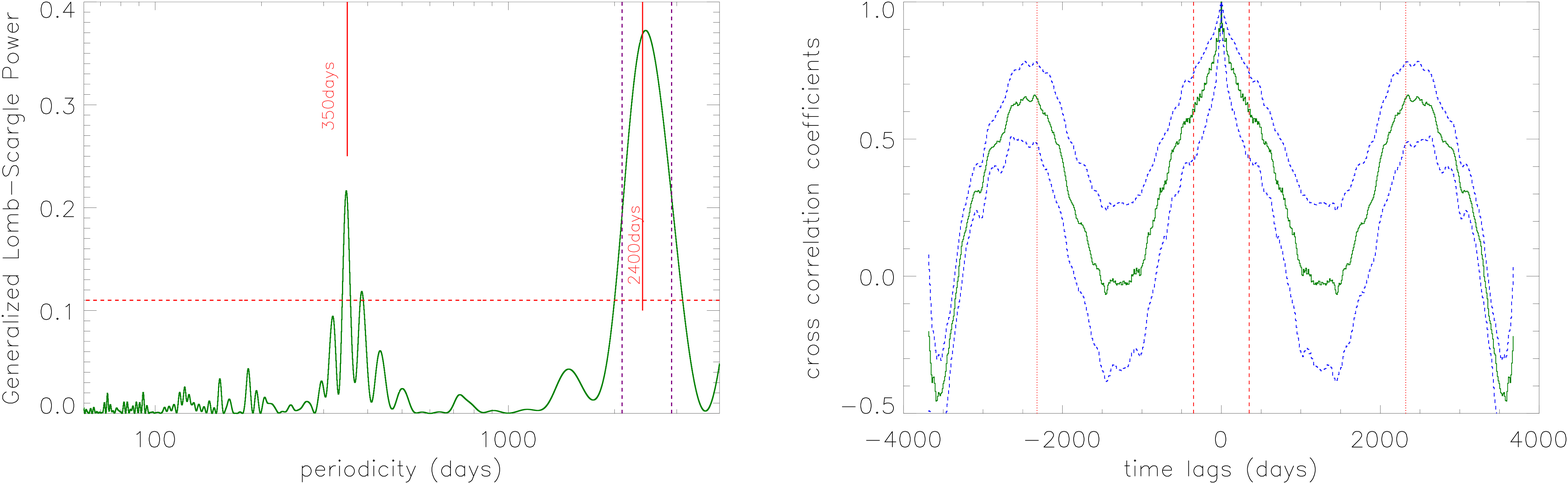}
\caption{Left panel shows results through the Generalized Lomb-Scargle periodogram. In the left panel, 
horizontal dashed red line shows the 99.99\% confidence level through the bootstrap method. 
The vertical red lines mark the two peaks around 350days and around 2400days. The vertical 
dashed purple lines (at 2100days and at 2900days) mark the periodicity width at the half maximum for 
the periodicity around 2400days. Right panel shows the ACF results. In the right panel, solid dark 
green line and dashed blue lines show the ACF results and the corresponding uncertainties determined 
by the bootstrap method. The vertical dashed and dotted red lines mark the positions with time lags at 
$\pm$350days and at $\pm$2320days, respectively.
}
\label{qpo}
\end{figure*}

\begin{figure}
\centering\includegraphics[width = 9cm,height=6.5cm]{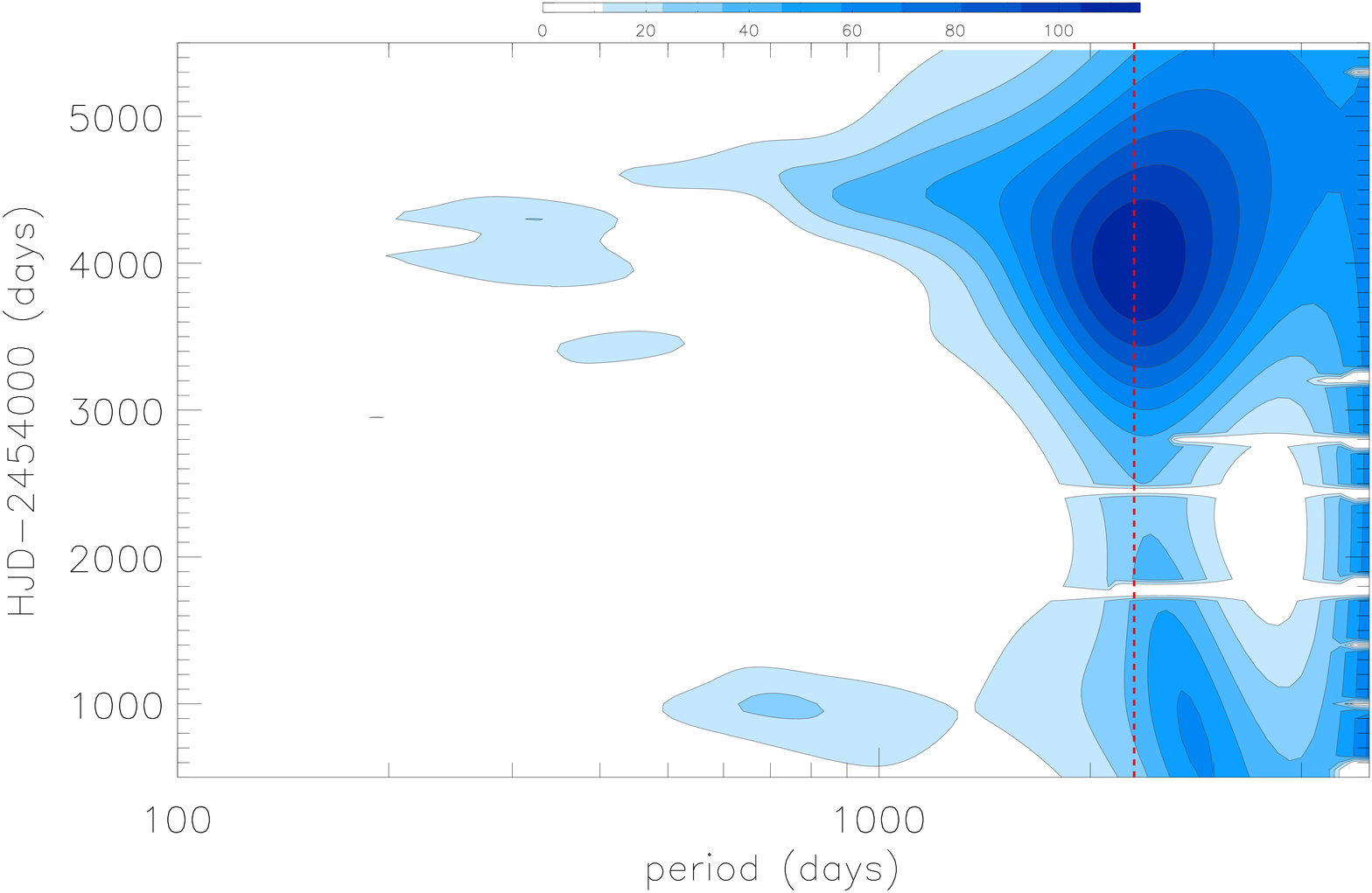}
\caption{The power maps in \obj~ determined by the WWWZ technique with frequency step of 0.00001 and 
searching periods from 100days to 5000days applied to the light curve shown in left panel of 
Figure~\ref{lmc}. The vertical red dashed line marks the position with periodicity about 2310days. 
}
\label{wwz}
\end{figure}

	Besides the direct fitting results shown in Figure~\ref{lmc} by the sinusoidal function, the 
improved Generalized Lomb-Scargle (GLS) periodogram \citep{ln76, sj82, zk09, vj18} is applied to check 
the periodicities in the long-term variabilities in SDSS J0752, similar as what have been discussed in 
\citet{zb16}. Higher than 99.99\% confidence level determined by the bootstrap method as 
discussed in \citep{ic19}, there are two periodicities around 350days and 2400days detected by the GLS 
periodogram shown in the left panel of Figure~\ref{qpo}. The GLS-determined periodicity about 2400days 
is well consistent with the determined 2321days shown in Figure~\ref{lmc}. Further discussions are 
shown on the GLS-determined periodicity around 350days by the following two points. First, accepted the 
periodicity around 350days, properties of the phase-folded light curve is re-checked. However, there 
are no sine-like variabilities in the phase-folded light curve with the periodicity around 350days. 
Second, the auto-correlation analysis (ACF) is applied to check the periodicities, shown in the right 
panel of Figure~\ref{qpo}. It is clear that only the time-lag around 2300days can lead to a clear peak 
in the cross correlation results, and no maximum peaks can be well detected around 350days in the ACF 
results. Moreover, QPOs in \obj~ is also determined through the commonly applied weighted wavelet 
z-transformation (WWZ) technique \citep{fg96} which have been well applied to determine QPOs as more 
recently discussed in \citet{al16, gt18, ks20, ly21}. The power maps determined by the WWZ technique are 
shown in Fig.~\ref{wwz} in \obj, leading to a clear peak around 2310days but no clear peaks around 
300days. Therefore, the periodicity around 2300days is well accepted, and the periodicity around 350days 
is not considered any more in the manuscript.


	Furthermore, the commonly accepted bootstrap method \citep{vj18} is applied to test robustness 
of the GLS-determined periodicity around 2400days in \obj. Among the shown data points in left panel 
of Figure~\ref{lmc}, about half of the data points are randomly collected to create a new light curve. 
Then, the GLS periodogram is applied to the new light curve to test whether the expected periodicity 
around 2400days has significance level lower than 99.99\%. Among 1000000 created light curves by randomly 
collected data points, there is no one new light curve has its GLS-determined periodicity with significance 
level lower than 99.99\%. Therefore, the false positive rate is less than $10^{-6}$ for the 
GLS-determined periodicity in left panel of Figure~\ref{qpo} in \obj.

	Finally, the QPOs with a periodicity around 2320days in SDSS J0752 can be well detected from 
the long-term light curve (time duration about 2.1 times longer then the detected periodicity) with 
confidence level higher than 99.99\%, based on the best-fitting results directly by the sinusoidal 
function shown in the left panel of Figure~\ref{lmc}, on the sine-like phase-folded light curve shown 
in the right panel of Figure~\ref{lmc}, on the results of GLS periodogram shown in the left panel of 
Figure~\ref{qpo}, on the ACF results shown in the right panel of Figure~\ref{qpo} and on the power maps 
determined by the WWZ technique shown in Figure~\ref{wwz}.

\begin{figure*}
\centering\includegraphics[width = 18cm,height=6cm]{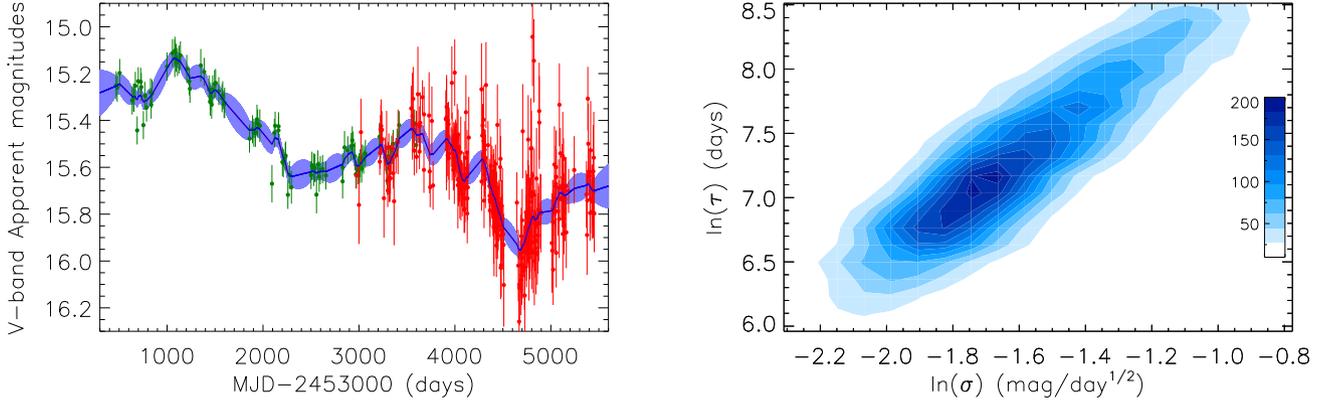}
\caption{Left panel shows the DRW-determined best descriptions to the long-term variabilities of 
SDSS J0752. Solid blue line and area filled with light blue show the best descriptions and the 
corresponding $1\sigma$ confidence bands. Symbols in dark green and in red represent the data values 
from the CSS and from the ASAS-SN. Right panel shows the MCMC determined two-dimensional posterior 
distributions of the DRW model parameters of $\ln(\tau)$ and $\ln(\sigma)$.
}
\label{drw}
\end{figure*}

\begin{figure}
\centering\includegraphics[width = 8cm,height=10cm]{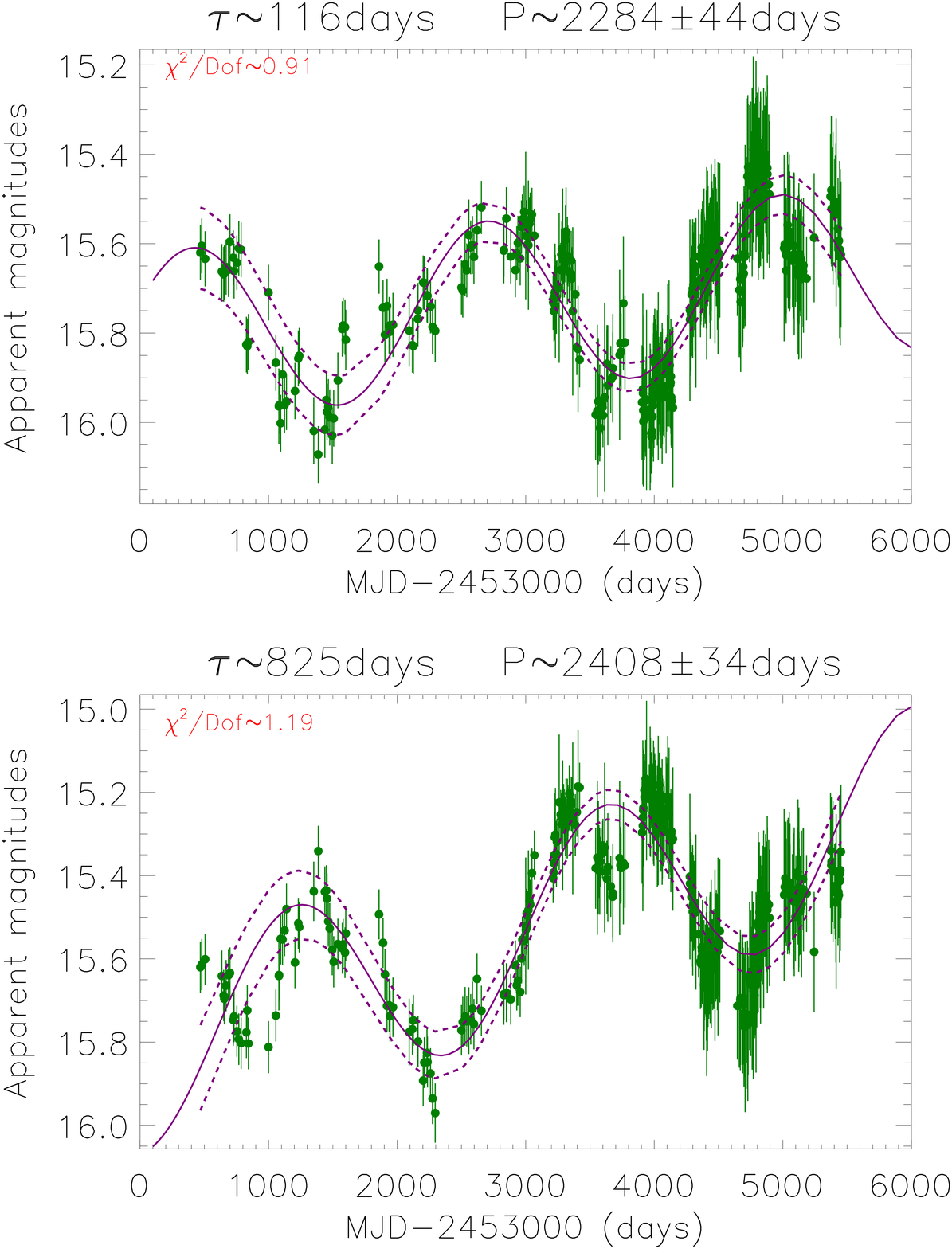}
\caption{Two examples on probable mis-detected QPOs in the simulating light curves by the CAR process. In 
each panel, solid dark green circles plus error bars show the simulating light curve, solid and dashed 
lines in purple show the best descriptions to the light curve and the corresponding 99.9999\% confidence 
bands, based on the sinusoidal function plus a linear trend. The input value of $\tau$ and the calculated 
periodicity $P$ are marked in the title of each panel. And the calculated $\chi^2/Dof$ is marked in the 
top left corner in each panel.
}
\label{fake}
\end{figure}

\subsection{Mis-detected QPOs in long-term AGN variabilities?}

	As more recently discussed in \citet{vu16}, false periodicities can be expected in quasar 
time-domain surveys. Therefore, it is necessary and interesting to determine whether the determined 
optical QPOs was mis-detected QPOs in the long-term optical variabilities of \obj, although the 
different techniques applied in the subsection above can lead to the detected optical periodicity 
in \obj. In the subsection, similar as the considered statistical method in \citet{mr16} to assess 
the significance of an absorption line in X-ray spectra, the following procedure is applied to 
estimate the probability to detected mis-detected QPOs in light curves well described by random 
walk process.

	It is well known that the CAR (or DRW) process is preferred to describe the intrinsic 
long-term AGN variabilities. Here, the DRW process is also applied to describe the long-term 
variabilities of SDSS J0752, through the public code JAVELIN (Just Another Vehicle for Estimating 
Lags In Nuclei) \citep{koz10, zk13} with two well accepted parameters of intrinsic characteristic 
variability amplitude and timescale of $\sigma$ and $\tau$. The best descriptions to the light curve 
are shown in the left panel of Figure~\ref{drw}. And the corresponding MCMC determined two dimensional 
posterior distributions of the parameters of $\sigma$ and $\tau$ are shown in the right panel of 
Figure~\ref{drw}, with the determined $\ln(\tau/days)\sim7.23\pm0.57$ ($\tau\sim1385$days) and 
$\ln(\sigma/(mag/dyas^{1/2}))\sim-1.64\pm0.27$ ($\sigma\sim0.19{\rm mag/day^{1/2}}$). Comparing with 
the long-term variabilities of SDSS quasars shown in Figure~3\ in \citet{mi10}, the DRW determined 
$\tau\sim1385{\rm days}$ is apparently one magnitude longer than the mean value around 200days of the 
SDSS quasars, and the DRW determined $\log(SF_{\infty}/mag)=\log(\sigma\times\sqrt(\tau))\sim0.86$ is 
definitely one magnitude larger than the mean value around -0.7 of the SDSS quasars. The larger 
values of $\sigma$ and $\tau$ than the common values of the SDSS quasars indicate SDSS J0752 is 
an interesting target.

	Based on the CAR process discussed in \citet{kbs09}:
\begin{equation}
\dif LMC_t=\frac{-1}{\tau}LMC_t\dif t+\sigma\sqrt{\dif t}\epsilon(t)~+~15.62
\end{equation}
where $\epsilon(t)$ a white noise process with zero mean and variance equal to 1, it is interesting 
to check whether CAR process determined long-term variability can lead to mis-detected QPOs. Here, 
the mean value of $LMC_t$ is set to be 15.62 (the mean value of the observational light curve of 
SDSS J0752), which has no effects on the following results. Then, a series of 10000 
simulating light curves [$t_i$, $LMC_i$] are created, with randomly selected values of $\tau$ from 
50days to 1000days and the $\sigma_*$ (the parameter in unit of mag in the CAR process in \citet{kbs09} 
slightly different from the $\sigma$ applied in the JAVELIN) leading the variance $\tau\sigma_*^2/2$ 
to be 0.047 (the variance of the observational light curve of SDSS J0752), and $t_i$ are the same 
as the observational time information shown in Figure~\ref{lmc}. And the similar uncertainties 
$LMC_{t,~err}$ are simply added to the simulating light curves $LMC_t$ by
\begin{equation}
LMC_{t,~err}~=~LMC_{t}\times\frac{L_{err}}{L_{obs}}
\end{equation}
with $L_{obs}$ and $L_{err}$ as the observational light curve and the corresponding uncertainties 
shown in left panel of Figure~\ref{lmc}.

	Then, the following two criteria are applied to check whether QPOs can be detected in the 
simulating light curves. First, GLS-determined periodicities should be around 2300days (larger 
than 2100days and smaller than 2900days, estimated from the corresponding periodicity width at the 
half maximum marked by vertical dashed purple lines in left panel of Figure~\ref{qpo}) with 
significance level higher than 99.99\%. Second, periodicities determined by the ACF results should 
be around 2300days (larger than 2100days and smaller than 2900days) with cross correlation coefficients 
higher than 0.5. Finally, among the 10000 simulating light curves, there are 8 light curves with 
expected mis-detected QPOs, accepted the two criteria above. Moreover, Figure~\ref{fake} shows 2 
of the 8 simulating light curves with mis-detected QPOs and the corresponding best-fitting results 
by the sinusoidal function. The results indicate that the DRW process (or the CAR process) can lead 
to light curves with mathematical determined QPOs (the mis-detected QPOs, or the fake QPOs), however, 
the probability of the mis-detected QPOs in CAR-process simulating light curves is around 0.08\% 
(8/10000). The results strongly indicate that the probability higher than 99.92\% (1-0.08\%) to 
support that the detected optical QPOs in SDSS J0752 are not mis-detected QPOs from a pure DRW 
process described light curve.

\begin{figure}
\centering\includegraphics[width = 8cm,height=5cm]{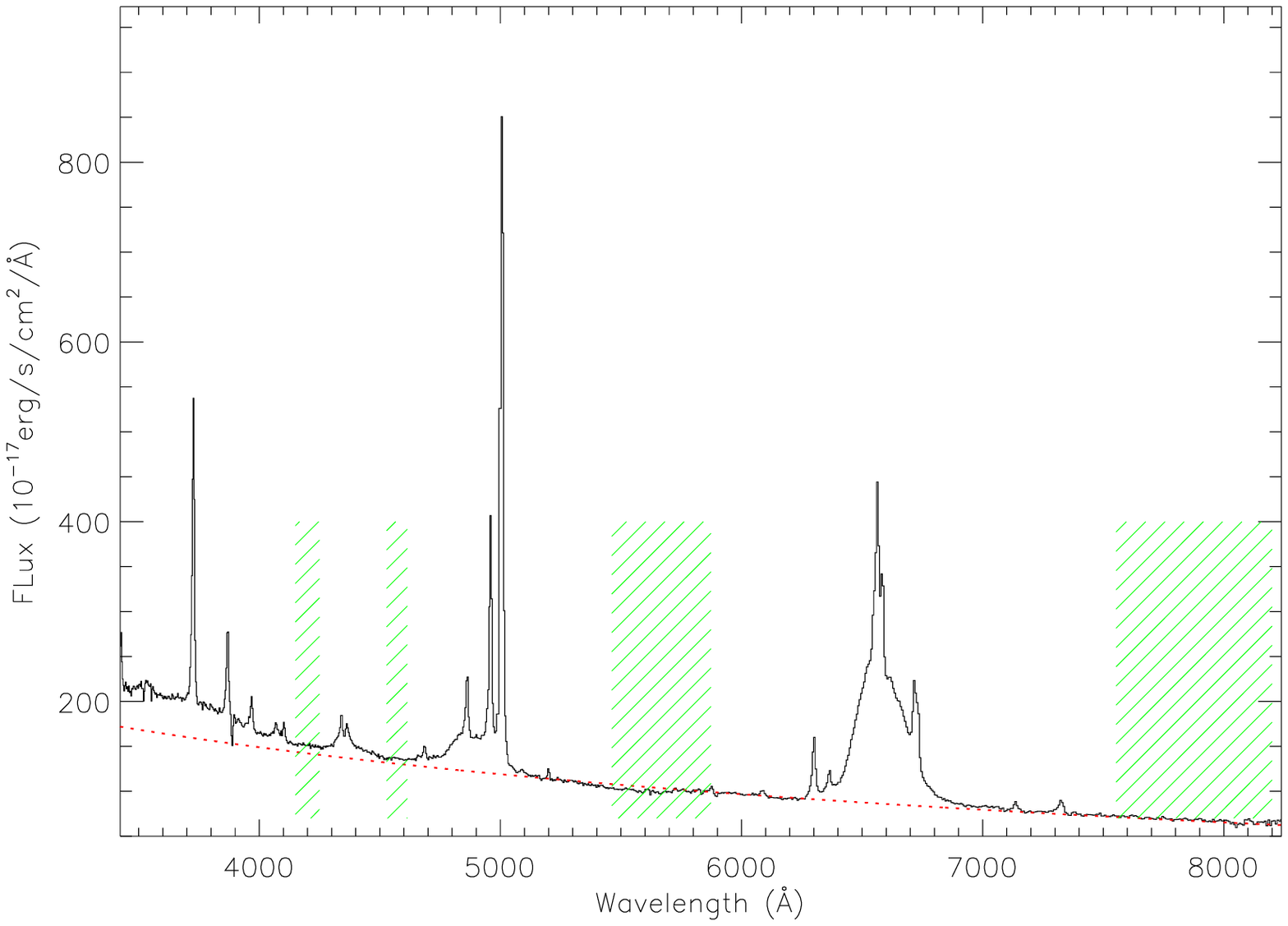}
\caption{The galactic reddening corrected spectrum of SDSS J0752\ in rest frame. The dotted red 
line represents the determined power-law continuum emissions. The areas filled with green lines 
show the wavelength windows applied to determine the power law continuum emissions.
}
\label{spec}
\end{figure}

\begin{figure*}
\centering\includegraphics[width = 8.5cm,height=5.5cm]{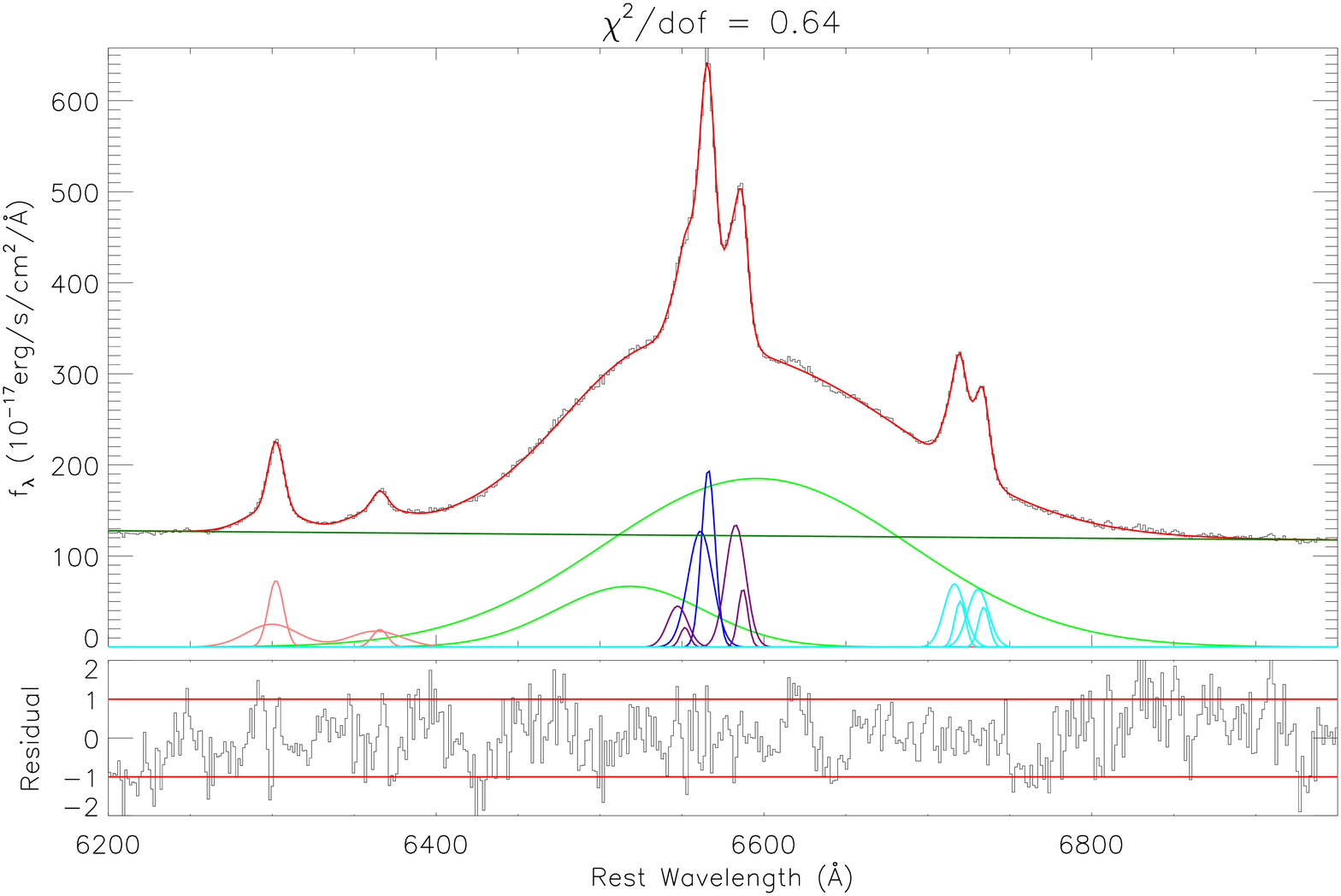}
\centering\includegraphics[width = 8.5cm,height=5.5cm]{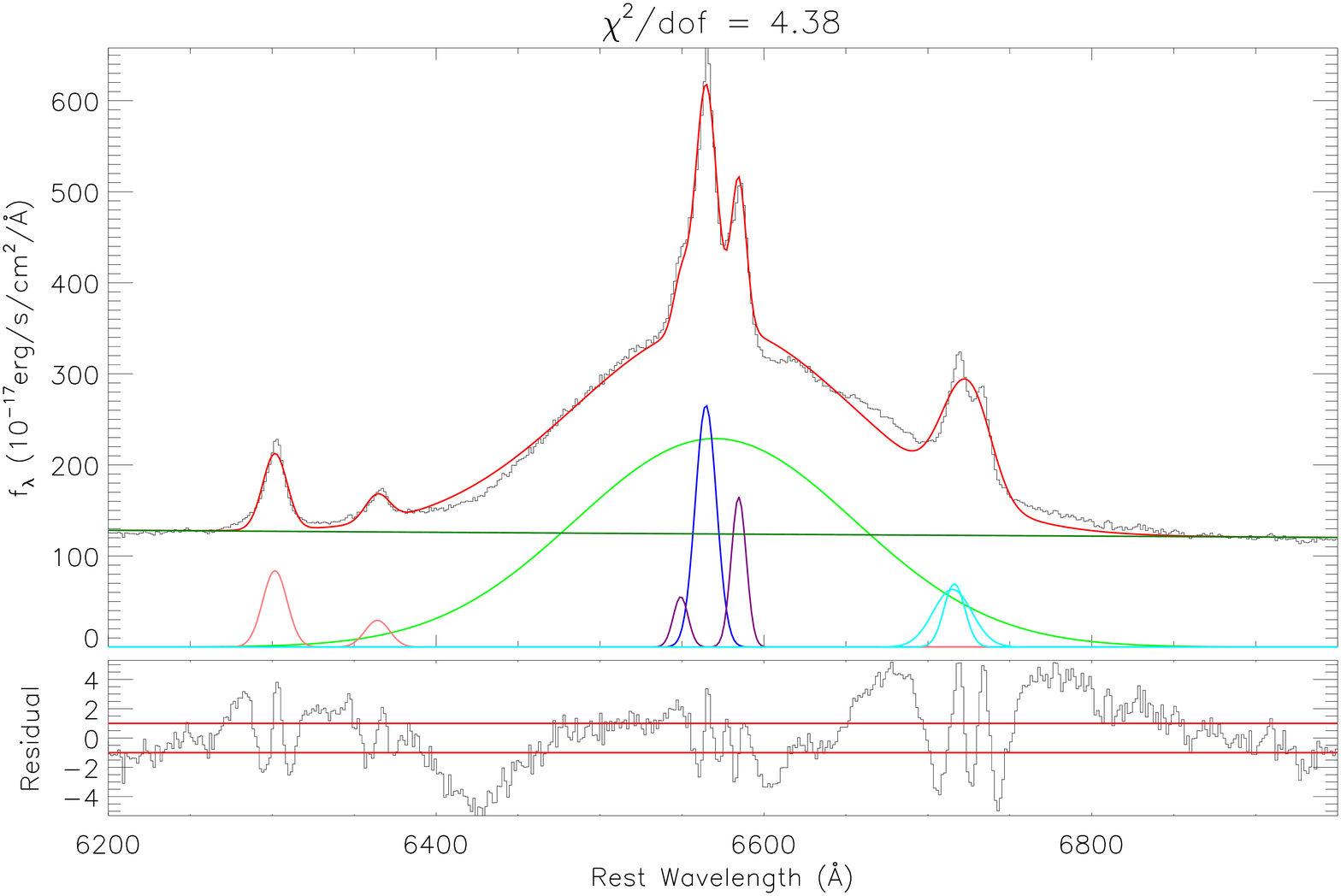}
\caption{Left panel shows the best-fitting results (top region) and the corresponding residuals 
(bottom region) to the emission lines around the H$\alpha$, by two broad Gaussian components in broad 
H$\alpha$, a core and an extended components in each narrow emissions line. In top region, solid black 
line shows the observed SDSS spectrum, solid red line shows the best-fitting results, solid dark green 
line shows the determined power law continuum emissions, solid green lines show the determined two broad 
components in the broad H$\alpha$, solid blue lines show the determine core and extended components in 
the narrow H$\alpha$, solid pink lines show the determined core and extended components in the [O~{\sc i}] 
doublet, solid purple lines show the determine core and extended components in the [N~{\sc ii}] doublet, 
and solid cyan lines show the determined core and extended components in the [S~{\sc ii}] doublet. In 
the bottom region, solid red lines show $Residual~=~\pm1$. Right panel shows the best-fitting 
results (top region) and the corresponding residuals (bottom region) to the emission lines around the 
H$\alpha$, by one broad Gaussian component in broad H$\alpha$, one Gaussian component in each narrow 
emissions line. In top region, solid black line shows the observed SDSS spectrum, solid red line shows 
the best-fitting results, solid dark green line shows the determined power law continuum emissions, solid 
green line shows the determined broad component in H$\alpha$, solid blue line shows the determined narrow 
component in H$\alpha$, solid purple lines show the determined [N~{\sc ii}] doublet, solid cyan lines 
show the determined [S~{\sc ii}] doublet, and solid pink lines show the determined [O~{\sc i}] doublet. 
In the bottom region, solid red lines show $Residual~=~\pm1$.
}
\label{ha}
\end{figure*}

\begin{table}
\caption{Line parameters}
\begin{tabular}{lllll}
\hline\hline
line &  & $\lambda_0$ & $\sigma$ & flux \\
\hline\hline
\multirow{2}{*}{Broad H$\alpha$}  &    &  6517.7$\pm$2.4  & 45.1$\pm$2.6  & 73$\pm$9 \\
	&  & 6594.9$\pm$1.6 & 93.8$\pm$0.6 & 440$\pm$8 \\
\hline
\multirow{2}{*}{Broad H$\beta$} &  & 4827.9$\pm$1.9 & 33.4$\pm$2.1 & 14$\pm$2 \\
	&  & 4885.1$\pm$1.1 & 69.5$\pm$0.4 & 82$\pm$3 \\
\hline
\multirow{2}{*}{[O~{\sc iii}]$\lambda5007$\AA} &
	c & 5008.4$\pm$0.1 & 3.9$\pm$0.1 &  84$\pm$2 \\
	& e & 5004.2$\pm$0.1 & 11.6$\pm$0.2  &  59$\pm$2 \\
\hline
\multirow{2}{*}{Narrow H$\beta$} & c & 4863.6$\pm$0.3 & 3.0$\pm$0.3 & 5.1$\pm$1.2 \\
	&  e & 4859.7$\pm$1.2 & 6.2$\pm$0.6 &  5.6$\pm$1.3 \\
\hline
\multirow{2}{*}{Narrow H$\alpha$} & c & 6566.2$\pm$0.3 & 3.9$\pm$0.2 & 18$\pm$2 \\
	& e & 6561.4$\pm$0.2 & 7.2$\pm$0.4 &  25$\pm$2 \\
\hline
\multirow{2}{*}{[N~{\sc ii}]$\lambda6583$\AA} &
c & 6586.8$\pm$0.2 & 3.4$\pm$0.4 & 6.7$\pm$1.3 \\
&  e & 6582.1$\pm$0.2 & 6.4$\pm$0.5 &  19.5$\pm$1.5 \\
\hline
\multirow{2}{*}{[O~{\sc i}]$\lambda6300$\AA} &
c & 6302.2$\pm$0.2 & 4.7$\pm$0.2 &  8.7$\pm$0.4 \\
& e & 6299.5$\pm$0.7 & 16.8$\pm$1.1 & 10.7$\pm$0.6 \\
\hline
\multirow{2}{*}{[O~{\sc i}]$\lambda6363$\AA} &
	c & 6365.7$\pm$0.2 & 4.8$\pm$0.2  & 2.4$\pm$0.2 \\
	& e & 6362.9$\pm$0.6 & 16.9$\pm$1.1  & 7.6$\pm$0.6 \\
\hline
\multirow{2}{*}{[S~{\sc ii}]$\lambda6716$\AA} &
c & 6719.6$\pm$0.3 & 3.3$\pm$0.5 &  4.2$\pm$1.7 \\
& e & 6716.2$\pm$0.7 & 6.4$\pm$0.4  & 11.1$\pm$1.9 \\
\hline
\multirow{2}{*}{[S~{\sc ii}]$\lambda6731$\AA} &
c & 6734.1$\pm$0.3 & 3.4$\pm$0.5 & 3.7$\pm$1.5 \\
& e & 6730.6$\pm$0.6 & 6.4$\pm$0.4  & 10.2$\pm$1.5 \\
\hline\hline
\end{tabular}\\
Notice: The first column shows which line is measured. The second column shows which component is determined 
in the narrow emission lines, 'c' means the core component, 'e' means the extended component. The third, fourth 
and fifth columns show the measured line parameters: the center wavelength $\lambda_0$ in unit of \AA, the line 
width (second moment) $\sigma$ in unit of \AA~ and the line flux in unit of ${\rm 10^{-15}~erg/s/cm^2}$. \\ 
For the broad H$\alpha$ and broad H$\beta$, there are two Gaussian components.  
\end{table}

\begin{figure}
\centering\includegraphics[width = 8cm,height=5.5cm]{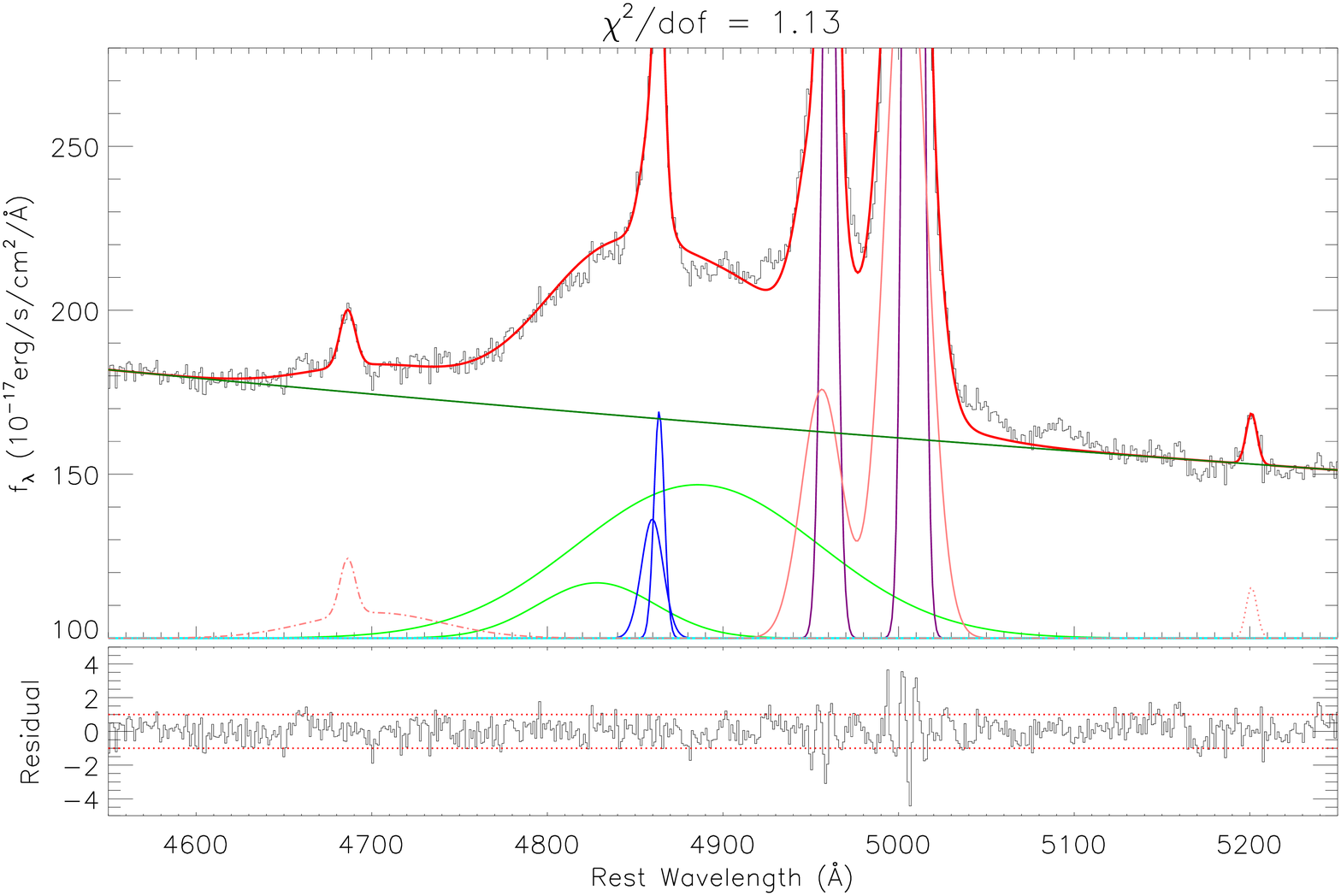}
\caption{Best-fitting results (top region) and the corresponding residuals (bottom region) to the 
emission lines around the H$\beta$. In top region, solid black line shows the observed SDSS spectrum around 
H$\beta$, solid red line shows the best-fitting results, solid dark green line shows the determined power law 
continuum emissions, solid green lines show the determined two broad components in the broad H$\beta$, solid 
blue lines show the determine core and extended components in the narrow H$\beta$, solid lines in purple and 
in pink show the determine core and extended components in the [O~{\sc iii}] doublet, dot-dashed pink line 
shows the determined broad and narrow component of He~{\sc ii} line, and dotted pink line shows the determined 
[N~{\sc i}]$\lambda$5205\AA~ line. Here, in top region, the plots are shown with $f_\lambda$ smaller than 
${\rm 200\times10^{-17}~erg/s/cm^2/textsc{\AA}}$, in order to clearly show the broad components in the H$\beta$. 
In the bottom region, solid red lines show $Residual~=~\pm1$.
}
\label{hb}
\end{figure}

\section{Spectroscopic properties of SDSS J0752}

	Figure~\ref{spec} shows the high-quality galactic reddening corrected (E(B-V)~=~0.0448) spectrum 
of SDSS J0752 with PLATE-MJD-FIBERID=1582-52939-0612. The apparently blue continuum emissions lead SDSS 
J0752 to be well classified as a SDSS quasar. The featureless continuum emissions can be well described 
by a power law function, $f_\lambda~\propto~\lambda^{-0.51}$ through the following four windows 
[4150\AA,~4250\AA], [4528\AA,~4614\AA], [5462\AA,~5873\AA] and [7553\AA,~8200\AA], leading the continuum 
luminosity at rest wavelength 5100\AA~ to be about $\lambda~L_{\rm 5100}~=~2.12~\times~10^{44}~{\rm erg/s}$.

	Then, emission lines around H$\alpha$ in SDSS J0752 can be well measured, similar as what we 
have recently done in \citet{zh21b, zh21c}. Left panel of Figure~\ref{ha} shows the emission lines within 
rest wavelength range from 6200\AA~ to 6850\AA, including [O~{\sc i}]$\lambda6300,~6363$\AA~ doublet, 
[N~{\sc ii}]$\lambda6548,~6583$\AA~ doublet, broad and narrow H$\alpha$, and [S~{\sc ii}]$\lambda6716,~6731$\AA~ 
doublet. The following model functions are applied to describe the emission lines, in order to obtain further 
clear properties of the broad components in the H$\alpha$. There are two broad Gaussian functions applied 
to describe the broad H$\alpha$, two Gaussian functions applied to describe each narrow emission line 
(a core component plus an extended component, similar as discussed in [O~{\sc iii}] 
doublet in \citet{gh05a, sh11, zh21b}), and a power law function applied to describe the AGN continuum 
emissions underneath the emission lines. Through the Levenberg-Marquardt least-squares minimization 
technique, the emission lines around H$\alpha$ can be well fitted. When the fitting procedure 
is running, only two restrictions are accepted. On the one hand, the flux ratios of components in 
[N~{\sc ii}] doublet are set to the theoretical values of 3. On the other hand, the emission flux of each 
Gaussian emission component is not smaller than zero. The best-fitting results $Y_{fit}$ and the 
corresponding residuals $(Y-Y_{fit})/Y_{err}$ (where $Y$ and $Y_{err}$ represents the SDSS spectrum 
and the corresponding uncertainties) are shown in Figure~\ref{ha}, with $\chi^2/dof~\sim~0.64$. 
The measured line parameters are listed in Table~1.

	Moreover, in order to confirm the extended components in narrow emission lines around H$\alpha$ 
and the two broad Gaussian components in broad H$\alpha$, the other one model is considered by including 
the following model functions. There is only one broad Gaussian function applied to describe the broad 
H$\alpha$, one Gaussian function applied to describe each narrow emission line. Then, through the same 
Levenberg-Marquardt least-squares minimization technique, the best fitting results and corresponding 
residuals are shown in the right panel of Figure~\ref{ha} with $\chi^2/dof~\sim~4.38$. Therefore, it is 
necessary and reasonable to consider the two broad Gaussian components in H$\alpha$ and the extended 
components in the narrow emission lines.

	Meanwhile, Figure~\ref{hb} shows the best fitting results (with $\chi^2/dof~\sim~1.13$) to the 
emission lines around H$\beta$ with rest wavelength from 4450\AA~ to 5350\AA~ by the following model 
functions through the same Levenberg-Marquardt least-squares minimization technique, to support the 
similar line profiles between broad H$\alpha$ and broad H$\beta$. There are two broad Gaussian components 
applied to describe the broad H$\beta$, four Gaussian components applied to describe the core and 
extended components in [O~{\sc iii}] doublet, two Gaussian components applied to describe the core 
and the extended component in narrow H$\beta$, one Gaussian component applied to describe the narrow 
[N~{\sc i}]$\lambda5205$\AA, two Gaussian functions applied to describe the broad and narrow He~{\sc i}, 
and a power law component applied to describe the continuum emissions underneath the emission lines. 
When the fitting procedure is running, the redshift and the line width of the two broad components in 
the broad H$\beta$ are the same as those of the two broad components in the broad H$\alpha$, and the 
flux ratios of the components of the [O~{\sc iii}] doublet are set to the theoretical values of 3. The 
measured line parameters of the emission lines around H$\beta$ are also listed in Table 1. 

	It is clear that the broad H$\alpha$ includes two broad components with line parameters of 
[$\lambda_0$, $\sigma_l$, $f$] (rest central wavelength in unit of \AA, second moment in unit of \AA~ 
and line flux in unit of $10^{-15}{\rm erg/s/cm^2}$) as [6517.7$\pm$2.4, 45.1$\pm$2.6, 73$\pm$9] and 
[6594.9$\pm$1.6, 93.8$\pm$0.6, 440$\pm$8]. In order to confirm the two broad components, new model 
functions with only one broad Gaussian component to describe the broad H$\alpha$ but two components 
(core and extended) to describe each narrow emission line are applied to re-describe the emission 
lines around the H$\alpha$, leading to $\chi^2/dof~\sim~3.03$, strongly indicating that the two broad 
components are preferred in the broad H$\alpha$: one blue-shifted component with shifted velocity about 
$2168\pm110{\rm km/s}$ and one red-shifted component with shifted velocity about $1384\pm74{\rm km/s}$, 
based on the central wavelength difference between the broad component and the core component of the 
narrow H$\alpha$ with line parameters listed in Table~1. Here, the best fitting results 
and corresponding residuals are not shown for the model functions with one broad Gaussian component 
applied to describe the broad H$\alpha$ and two components applied to describe each narrow emission 
line, due to similar residuals with rest wavelength from 6400\AA~ to 6800\AA~ shown in the right panel 
of Figure~\ref{ha}.

\section{Main Discussions}

	The determined two broad H$\alpha$ components provide better chances to check whether is 
there a central BBH system in SDSS J0752, because the two BH masses in the probable BBH system can 
be well determined by properties of the two broad components. Accepted assumption of a central BBH 
system in SDSS J0752, through the virialization assumption \citep{pe04, vp06} to broad line emission 
clouds combining with the empirical R-L relation \citet{bd13} to determine the distance between 
emission clouds and central BH, the virial BH mass can be estimated through the formula discussed 
in \citet{gh05}
\begin{equation}
\frac{M_{\rm BH}}{\rm M_\odot}=2.2\times10^6\times(\frac{L_{\rm H\alpha}}{\rm 10^{42}erg/s})^{0.56}
	\times(\frac{FWHM_{\rm H\alpha}}{\rm 1000km/s})^{2.06}
\end{equation}.
Here, the more recent R-L relation discussed in \citet{bd13} is applied. Then, based on the line 
luminosities and line widths of the two broad Gaussian components, the two BH masses can be estimated 
as $M_{BH,b}=(8.8\pm1.7)\times10^7{\rm M_\odot}$ for the blue-shifted BH system and 
$M_{BH,r}=(104.4\pm3.1)\times10^7{\rm M_\odot}$ for the red-shifted BH system. Then, following the 
discussed results in \citet{eb12}, the space separation between the central two BHs can be estimated as 
\begin{equation}
A_{BBH}=0.432\times M_{8}\times(\frac{P_{BBH}/year}{2652M_{8}})^{2/3}\sim0.02pc
\end{equation}
where $M_{8}$ represent the total BH mass of the BBH system in unit of $10^8{\rm M_\odot}$ and 
$P_{BBH}\sim6.4{\rm yr}$ represents the orbital period of the BBH system.

	Meanwhile, besides the BBH system, precessions of emission regions with probable hot spots 
for the optical continuum emissions can also be applied to describe the detected optical QPOs in 
SDSS J0752. Not considering two independent broad H$\alpha$ components, line properties of the total 
broad H$\alpha$ can be applied to estimate the central virial BH mass as $1.87~\times~10^9~{\rm M_{\odot}}$, 
through the formula discussed in \citet{pe04} 
\begin{equation}
M_{\rm BH}~=~5.5~\times~ \frac{\sigma^2_{\rm broad, H\alpha}~\times~ R_{\rm BLRs}}{G}
\end{equation}
where $\sigma_{\rm broad, H\alpha}\sim4010~{\rm km/s}$ represents the second moment of the total broad 
H$\alpha$ and $R_{\rm BLRs}\sim54$ light-days is the distance of broad line emission regions to central 
BH estimated through the R-L relation \citep{bd13} with the continuum luminosity at 5100\AA~ about 
$2.12~\times~10^{44}~{\rm erg/s}$ shown in Figure~\ref{spec}. Then, as discussed in \citet{eh95} and 
in \citet{st03}, the expected disk precession period can be estimated as 
\begin{equation}
	T_{\rm pre}\sim1040M_{8}R_{3}^{2.5}yr
\end{equation},
where $R_{3}$ and $M_{8}\sim18.7$ mean the distance of emission regions to central BH in unit of 
$10^3R_{G}$ ($R_G=\frac{GM_{BH}}{c^2}$) and the BH mass in unit of $10^8{\rm M_\odot}$. In order 
to get the detected periodicity about 6.4yr, the expected $R_3$ could be around 0.041 (41$R_G$). However, 
based on the discussed results on the distance of NUV emission regions to central BHs in \citet{mc10} 
through the microlensing variability properties of eleven gravitationally lensed quasars, the NUV 
2500\AA~ continuum emission regions in SDSS J0752 have distance from central BH as 
\begin{equation}
\begin{split}
	\log{\frac{R_{2500}}{cm}}=&15.78+0.80\log(\frac{M_{BH}}{10^9M_\odot})\\
	&\sim(72\pm20)R_G
\end{split}
\end{equation}
The estimated NUV emission regions have distances about two times larger than the optical continuum 
emission regions in SDSS J0752 under the disk precession assumption. The unreasonable results 
strongly indicate that the disk precessions of emission regions are not preferred to be applied to 
explain the detected optical QPOs in SDSS J0752.

	Moreover, an oscillating recoiled supermassive black hole (rSMBH) scenario has been well 
discussed in \citet{km18}, and the rSMBH scenario has been applied to explain the observed velocity 
offsets of the blue/red-shifted components in asymmetric broad emission lines in Mrk1018. Therefore, 
it is interesting to consider whether the rSMBH scenario can be applied to explain the blue/red-shifted 
components in broad Balmer lines in \obj. As discussed in \citet{km18}, the rSMBH scenario can lead 
to constant velocity offset difference between blue-shifted and red-shifted broad components, however, 
BBHs system should lead to time dependent velocity offset difference between blue-shifted and red-shifted 
broad components. However, due to single-epoch spectra of \obj, properties of velocity offset difference 
can not be applied to determine whether the rSMBH scenario is preferred or not. However, \citet{km18} 
also discussed that a BBH system can indicate that a broad emission component with larger velocity 
offset will have a smaller line width at each epoch. In broad Balmer lines of \obj, the blue-shifted 
component has velocity offset 47\AA~ and line width 45.1\AA, but the red-shifted component has velocity 
offset 31\AA~ and line width 93.8\AA, consistent with the expected results by a BBH system: the 
blue-shifted component has larger velocity offset but smaller line width. Therefore, in the current 
stage, the BBH system is preferred in \obj, unless there are long enough spectroscopic variabilities 
of broad emission lines in \obj.

	As discussed in Introduction, long-term QPOs can be detected in blazars due to jet precessions. 
However, radio power at 1.4GHz is about 25mJy in SDSS J0752, provided by the NVSS (NRAO VLA Sky Survey, 
\citet{oi06}). Combining with the optical luminosity at 5100\AA, the radio loudness of SDSS J0752 is 
about $R~\sim~0.28$, indicating SDSS J0752 is a radio quiet AGN. Therefore, jet precessions can be well 
ruled out to explain the optical QPOs in SDSS J0752.

\section{Summaries and Conclusions}
    The main conclusions are as follows. 
\begin{itemize}
\item The long-term light curve from the CSS and the ASAS-SN can be well described by a sinusoidal function 
with a periodicity about 2321days (6.4yr) in SDSS J0752, which can be further confirmed by the corresponding 
sine-like phase folded light curve.
\item The periodicity can be re-confirmed by the Generalized Lomb-Scargle periodogram with confidence 
level higher than 99.99\%, and by the auto-correlation analysis results, and by the WWZ technique. 	
\item The long-term optical QPOs can be applied as the better indicators to the central BBH system in 
SDSS J0752. Based on the two broad H$\alpha$ components, the virial BH masses of the two central BHs 
can be estimated as $8.8\times10^7{\rm M_\odot}$ and $1.04\times10^9{\rm M_\odot}$, leading the BBH 
system expected space separation is about 0.02pc.
\item Based on the estimated sizes about $70R_G$ of the NUV emission regions about two times larger than 
the disk precession expected sizes about $41R_G$ of the optical emission regions, the disk precessions 
can be well ruled out to explain the detected QPOs in SDSS J0752. Meanwhile, the radio loudness about 0.28 
strongly support that the jet precessions can be totally ruled out to explain the detected QPOs in SDSS J0752.
\item Based on the mathematical CAR process simulating light curves as the intrinsic AGN variabilities, 
0.08\% probability can be determined to detect mis-detected QPOs in the simulating light curves. 
Therefore, the optical QPOs in SDSS J0752 are more confident. 
\end{itemize}

\section*{Acknowledgements}
Zhang gratefully acknowledge the anonymous referee for giving us constructive comments and 
suggestions to greatly improve our paper. Zhang gratefully acknowledges the kind support of Starting 
Research Fund of Nanjing Normal University, and the kind grant support from NSFC-12173020. This paper 
has made use of the data from the SDSS projects. The SDSS-III web site is \url{http://www.sdss3.org/}. 
SDSS-III is managed by the Astrophysical Research Consortium for the Participating Institutions of the 
SDSS-III Collaboration. This paper has made use of the data from the CSS projects 
\url{http://nesssi.cacr.caltech.edu/DataRelease/} and the ASAS-SN projects \url{https://asas-sn.osu.edu/}.

\section*{Data Availability}
The data underlying this article will be shared on reasonable request to the corresponding author
(\href{mailto:xgzhang@njnu.edu.cn}{xgzhang@njnu.edu.cn}).

\label{lastpage}
\end{document}